\pdfoutput=1

\documentclass{egpubl}
\usepackage{eurovis2018}
\SpecialIssuePaper         %

 \electronicVersion %

\ifpdf \usepackage[pdftex]{graphicx} \pdfcompresslevel=9
\else \usepackage[dvips]{graphicx} \fi

\PrintedOrElectronic

\usepackage{t1enc,dfadobe}

\usepackage{egweblnk}
\usepackage{cite}

\usepackage{textcomp}
\usepackage{color,soul}
\usepackage[dvipsnames]{xcolor}
\usepackage{subfigure}
\usepackage[final]{changes}
\usepackage{todonotes}

\definecolor{noDizzy}{HTML}{FDD0A2}
\definecolor{dizzy}{HTML}{D94801}

\definecolor{fourth}{HTML}{BAE4B3}
\definecolor{third}{HTML}{74C476}
\definecolor{second}{HTML}{31A354}
\definecolor{first}{HTML}{006D2C}

\newcommand{\saveparinfos}{%
\edef\myindent{\the\parindent}%
\edef\myparskip{\the\parskip}}

\usepackage{multirow}

\title[Maps and Globes in Virtual Reality]%
      {Maps and Globes in Virtual Reality}

\author[Y. Yang et al.]
       {\parbox{\textwidth}
       		{\centering 
       			Yalong Yang$^{1,2}$, Bernhard Jenny$^{1}$, Tim Dwyer$^{1}$, Kim Marriott$^{1}$, Haohui Chen$^{2}$ and Maxime Cordeil$^{1}$
       		}
        \\
         \parbox{\textwidth}
         	{\centering 
         		$^1$Faculty of Information Technology, Monash University, Australia\\
         		$^2$Data61, CSIRO, Victoria, Australia
         	}
       }

\begin{document}
\teaser{\centering
\vspace{-3em}
\subfigure[Exocentric globe] {
\includegraphics[height=2.5cm]{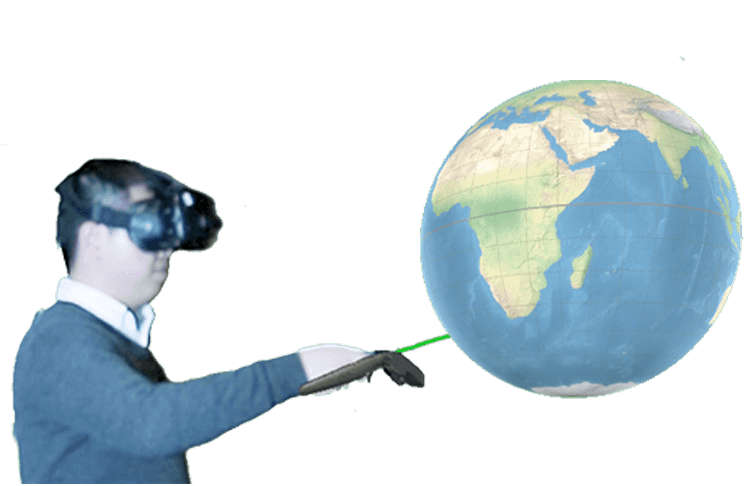}
\label{fig:exocentricglobeteaser}
}
\subfigure[Flat map] {
\includegraphics[height=2.5cm]{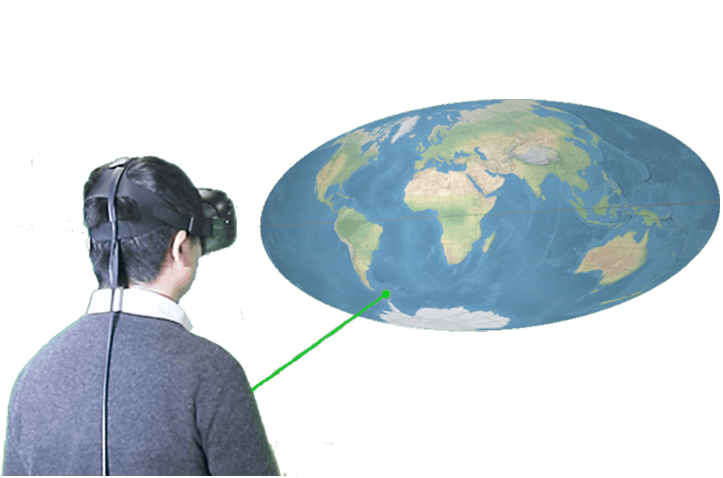}
\label{fig:flatmapteaser}
}
\subfigure[Egocentric globe] {
\includegraphics[height=2.5cm]{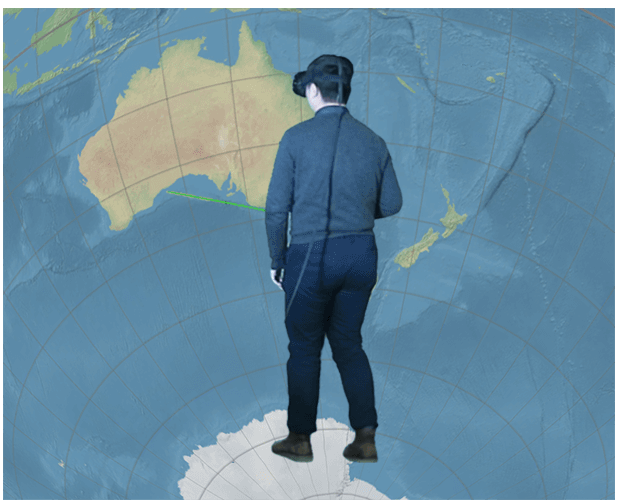}
\label{fig:egocentricglobeteaser}
}
\subfigure[Curved map] {
\includegraphics[height=2.5cm]{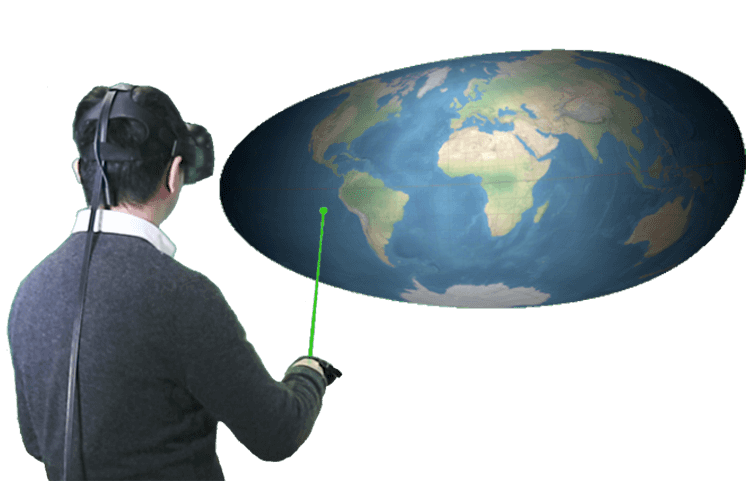}
\label{fig:curvedmapteaser}
}
 \caption{Four interactive visualisations for geographic data in Virtual Reality (VR).}
 \label{fig:teaser}
}
\maketitle

\begin{abstract}
This paper explores different ways to render world-wide geographic maps in virtual reality (VR).
We compare: (a) a 3D exocentric globe, where the user's viewpoint is outside the globe; (b) a flat map (rendered to a plane in VR); (c) an egocentric 3D globe, with the viewpoint inside the globe; and (d) a curved map, created by projecting the map onto a section of a sphere which curves around the user. In all four visualisations the geographic centre can be smoothly adjusted with a standard handheld VR controller and the user, through a head-tracked headset, can physically move around the visualisation.
For distance comparison \deleted{we find that the} exocentric globe is more accurate than egocentric globe and flat map.  For area comparison \deleted{we find}
more time is required with exocentric and egocentric globes than with flat and curved maps. For direction estimation, the exocentric globe is more accurate and faster than the other visual presentations. Our study participants had a weak preference for the exocentric globe. 
Generally the curved map had benefits over the flat map.  In almost all cases the egocentric globe was found to be the least effective visualisation.
Overall, our results provide support for the use of exocentric globes for geographic visualisation in mixed-reality.


\end{abstract}

\section{Introduction}
\vspace{-0.5em}
Maps and globes are widely used to visualise geographic data. They underpin how we understand the world and our place on it.  They are the foundation for geovisualisation and thematic cartography, in which qualitative and quantitative data with a spatial aspect is overlaid on its geographic location~\cite{slocum2009thematic}. Geovisualisation is widely used to understand both social and physical data, for instance  analysing census data \cite{Martin:1989ii}, studying epidemiology \cite{Moore:1999cc}, planning urban transportation policies \cite{Arampatzis:2004hw}, and exploring animal migration patterns \cite{Kolzsch:2013vq,Slingsby:2017vc}. 

\vspace{-0.2em}
Terrestrial maps and globes were used by the Ancient Greeks more than two thousand years ago. During the Renaissance matched pairs of celestial and terrestrial globes were the main tool for teaching geography and cosmology~\cite{dekker2007globes}, but in subsequent centuries the use of globes declined as they are more expensive to produce than maps, much bulkier to store, and do not scale. Additionally, from the Renaissance onwards many sophisticated map projections were invented~\cite{snyder1997flattening} that at least partly overcame the great disadvantage of a map: that it is not possible to draw the surface of the earth on a 2D surface without significant spatial distortion.

\vspace{-0.2em}
In the 21st century, however, the globe has made a remarkable comeback. Virtual globes have become familiar to most VR users, e.g.\ \cite{GoogleEarthVR}. 
    \added{With the arrival of commodity head-mounted displays (HMDs) for VR (e.g. HTC Vive) and also augmented- and mixed-reality (AR and MR, e.g. Microsoft Hololens, Meta2 and so on), we can expect to see more virtual globes being used in geographic visualisation applications. MR in particular, has great appeal in situated analytics scenarios \cite{elsayed2015situated} where visualisations are made available in challenging situations (in the field, surgery, or factory floor), and also for collaborative visualisation scenarios, where two or more people wearing HMDs can each see and interact with the globe, while still seeing each other directly for communication \cite{Cordeil:2016io}.}
A key question, however, is whether virtual globes are the best way to show global geographic data in immersive environments or whether maps or some other visualisations may be better. Surprisingly, given the fundamental importance of this question for the design of geovisualisation applications in VR and \replaced{MR}{mixed-reality}, it has not been formally tested previously. Our paper fills this gap by making two main contributions:

First, in Sec.~\ref{sec:methods} we present four different interactive visualisations of the earth's geography designed for use in VR for head mounted displays (HMD). The first two are well-known: the \textit{exocentric globe} (Fig.~\ref{fig:exocentricglobeteaser}) and the \textit{flat map} (Fig.~\ref{fig:flatmapteaser}). The other two are more novel. The standard globe is an \emph{exocentric} visualisation such that the viewer stands outside the globe.  An alternative approach is to place the viewer inside the globe~\cite{Zhang:2016bi}. With our \emph{egocentric globe} (Fig.~\ref{fig:egocentricglobeteaser}) the viewer sees a map projected onto a surrounding 360$^\circ$ sphere. One possible advantage of the egocentric sphere is that, if the viewpoint is close to the centre of the sphere, the inside surface of the globe is a constant distance away, reducing perceptual distortion. Our fourth visualisation is a novel VR visualisation we call the \emph{curved map} (Fig.~\ref{fig:curvedmapteaser}). This is a map projected onto a section of a sphere. The viewer faces the concave side of the map, so again a possible advantage is that the distance from the viewpoint to the surface of the map is relatively constant resulting in reduced perceptual distortion. 
All four visualisations support the same basic interaction, detailed in Section \ref{sec:interaction}.  The user can interactively move any geographic location to the centre of the view, an interaction we found to be essential to enable the tasks tested in our study. Users can also change their viewpoint through headtracked motion standard to modern VR.

Our second main contribution is a controlled study with HMD in VR investigating user preferences and the efficacy (accuracy and time) of these four different interactive visualisations (Sec.~\ref{sec:study}). We evaluate three fundamental spatial analysis tasks: distance and area comparison, as well as estimation of orientation between two locations.  We also analyze physical movement and user interaction with the visualisations.  

The ego- and exo-centric globes and curved maps are naturally 3D visualisations, while the flat map view is a 2D visualisation. 
Testing across different devices (e.g.\ flat map on regular screen versus 3D visualisations in VR) would introduce a large number of variables into the evaluation.  For example: resolution; head-tracked vs non-headtracked interaction; comfort of headset; and so on -- all of which are purely a function of the limitations of the current (rapidly developing) technology.  
Thus, we evaluate all four visualisations in VR such that these variables are eliminated.  Rather, we can focus on the geometry of the geographic surface over differences between devices. 

The results of our study (Sec.~\ref{sec:discussion}) show that the exocentric globe is the best choice for most of the tested tasks.  This surprised us, since in such a display only half of the geographic surface is visible to the user at any time.  Further, it has the most perceptual distortion (Sec.~\ref{sec:methods}), though less distortion due to projection than the flat and curved map. 
Though not as effective overall, for some tasks the flat map and curved map also perform well and were preferred by some participants.  
This result motivated a prototype implementation of a novel interaction for smoothly transitioning from exocentric globe to flat map and back (Sec.~\ref{sec:conclusion}).

\vspace{-1em}
\section{Related Work}
\noindent\textbf{\textit{Immersive visualisation with geospatial data:}}
Commodity VR and AR HMD has fuelled interest in so-called \emph{immersive analytics}, the exploration of emerging immersive interaction and display technologies to support analytical reasoning and decision making \cite{Chandler:2015eb}. 
Since most geospatial data is inherently three-dimensional, analysing geospatial data in VR is an obvious fit for immersive analytics.  Indeed the use of VR for geographic visualisation was an early application of VR \cite{heldley:2015fk}. Today, VR is used in various geospatial fields, such as architecture and environmental planning \cite{Portman:2015cb}. 
A recent review~\cite{dubel20142d}  provides a framework for understanding 2D and 3D presentation of spatial data, clarifying that the choice of 2D or 3D applies to both the \emph{reference space} and the \emph{attribute space}. Here we are concerned with 2D and 3D presentation of the reference space, i.e. the surface of the earth.
Exocentric globes can only show one hemisphere, which is a limitation for many global visualisations. GeoPeels was a proposal to create a hybrid between an exocentric globe and a flat map to address this limitation \cite{Debiasi:2015ul}. GeoPeels bends selected lobes of the hidden hemisphere into the field of view, resulting in a visualisation resembling a semi-peeled orange.

Kwon et al.~\cite{Kwon:2016go} introduced an egocentric spherical layout for networks in which the layout was wrapped around the head of the user in VR. They found that such immersive layouts were more effective for typical graph visualisation tasks (e.g.\ path and neighbour finding)  than traditional non-immersive 2D layouts. This idea was extended to geospatial data (for the visualisation and analysis of geospatial networks), resulting in immersive egocentric globes \cite{Zhang:2016bi}. However, the interaction design for egocentric globes has not been discussed, and no controlled user study has been conducted to evaluate their effectiveness in comparison to other geospatial VR visualisations.

\noindent\textbf{\textit{Map perception tasks for user studies:}}
A series of cognitive map studies have sought to better understand the creation, preservation and recall of geographic maps. 
Tasks tested include the estimation of area \cite{Battersby:2009jo} and position or distance (for example, \cite{Friedman:2000ux,Friedman:2006ux,Carbon:2010bo}). 
Recently, the continuity of paths  across edges of world maps has also been studied. It has become apparent that map edges have a negative impact on the perception of path continuity \cite{Hennerdal:2015do} and the estimation of distances \cite{Hruby:2016eu}. 

\noindent\textbf{\textit{Map projections and distortion:}}
A map projection transforms a sphere to a flat surface~\cite{Snyder:1987tk}. Hundreds of map projections have been devised~\cite{snyder1997flattening}, however, all map projections either distort angles or the relative size of areas. No projection can preserve the distances between all locations on a map and most only preserve distance along very few, carefully selected lines. Cartographers compromise in choice of projection based on the area displayed and the purpose of the map\cite{Jenny:2017it, Savric:2016do}.

\vspace{-1em}
\section{Showing the Earth in VR}
\label{sec:methods}
In this section we describe the four geospatial VR visualisations that we tested in the user study. We  compare the different visualisations in terms of the amount of distortion introduced by projecting the earth's surface onto the visualisation surface, the amount of perceptual distortion, and also the amount of the earth's surface that is visible to the viewer (Table~\ref{tab:comparison}). 
Perceptual distortion arises because the viewer essentially sees the VR image as a 2D image projected onto each eye and the visual system must reconstruct the position of elements on the surface using depth cues such as linear perspective, texture gradient, etc.~\cite{ware2012information}. The degree of perceptual distortion depends upon the relative position of the viewer and representation. It is also influenced by the depth cues provided by the VR environment. All visualisations were implemented in the Unity3D engine for the HTC Vive headset. The HTC Vive provides head-tracked stereoscopic VR and we used the Unity3D engine to provide linear perspective, texture gradient and shadows from a light source placed above and behind the viewer. 

\begin{table}
	\centering
	\footnotesize
	\begin{tabular}{l|l|l|l|}
\cline{2-4}
& \multicolumn{2}{l|}{Amount of distortion} & \multirow{2}{*}{\begin{tabular}[c]{@{}l@{}}Approximate area \\in field of view\end{tabular}} \\ \cline{2-3}
& Projection& Perceptual& \\ \hline
\multicolumn{1}{|l|}{Exocentric Globe} & None & High on edge & Hemisphere \\ \hline
\multicolumn{1}{|l|}{Flat Map}         & High & Medium       & Entire sphere \\ \hline
\multicolumn{1}{|l|}{Egocentric Globe} & None & Low          & Hemisphere \\ \hline
\multicolumn{1}{|l|}{Curved Map}       & High & Very low     & Entire sphere  \\ \hline
\end{tabular}
	\caption{Estimate of distortion and field of view of the four visualisations. Perceptual distortion and field of view are estimated for the initial position of the viewer.}
	\label{tab:comparison}
	\vspace{-2.5em}
\end{table}

\noindent\textbf{\textit{Exocentric Globe:}}
A direct three-dimensional rendering of the spherical model (Fig.~\ref{fig:exocentricglobeteaser}). 
Our exocentric globe has an invariable radius of 0.4 metres. The globe is initially positioned 1 metre in front of  the user at head height.
Because of occlusion the user can see at most one hemisphere. There is no distortion due to projecting the earth's surface onto the surface of the globe, but there is areal and angular distortion along the edges of the visible hemisphere.%

\noindent\textbf{\textit{Flat Map:}}
An elliptical projection is texture-mapped onto a quad measuring 1$\times$0.5 metres and is placed 1 metre from the user (Fig.~\ref{fig:flatmapteaser}). For vector data, we use our own partial port of the D$^3$ library for spherical rotation, cutting, clipping and resampling \cite{Bostock:2013ix}. For raster images we use a GPU rendering technique \cite{Jenny:2015hf}.

We chose the \emph{Hammer} map projection, which preserves the relative size of areas. 
To reduce the distortion of shapes and also for aesthetic reasons, we chose to use an equal-area projection with an elliptical boundary. We did not to use a projection that shows the poles as lines, when the map centre can be adjusted by the user as this is potentially confusing. 
We preferred the Hammer to the area-preserving Mollweide projection for computational efficiency. In comparison to other projections for world maps, the Hammer projection adds small distortion to distances in the central area of the map.

The great advantage of a map over a globe is that the entire surface of the earth is visible. While the Hammer map projection does not distort the relative size of areas, it introduces angular distortion, which increases away from the center of the map. There is some perceptual distortion because the distance between the viewer and the display surface varies; perceptual distortion increases as the viewer nears the map. Foreshortening distortion results when the map is viewed with an oblique angle.

\noindent\textbf{\textit{Egocentric Globe:}}
In a design (Fig.~\ref{fig:egocentricglobeteaser}) following Zhang et al. \cite{Zhang:2016bi}, we initially set the radius to 3 meters, and positioned the user at the centre of the globe.  However, pilot participants reported that they felt their field of view was limited when they are at the centre and wished to move to the edge of the globe to enlarge their field of view. So we adjusted the position of the globe to allow the participants to stand at the edge of the globe (80\% of radius away from the centre). To ensure that participants cannot walk out of the sphere, we increased the radius to 8 meters, which is larger than the walkable space.

\begin{figure}
	\centering
	\includegraphics[width=0.43\textwidth]{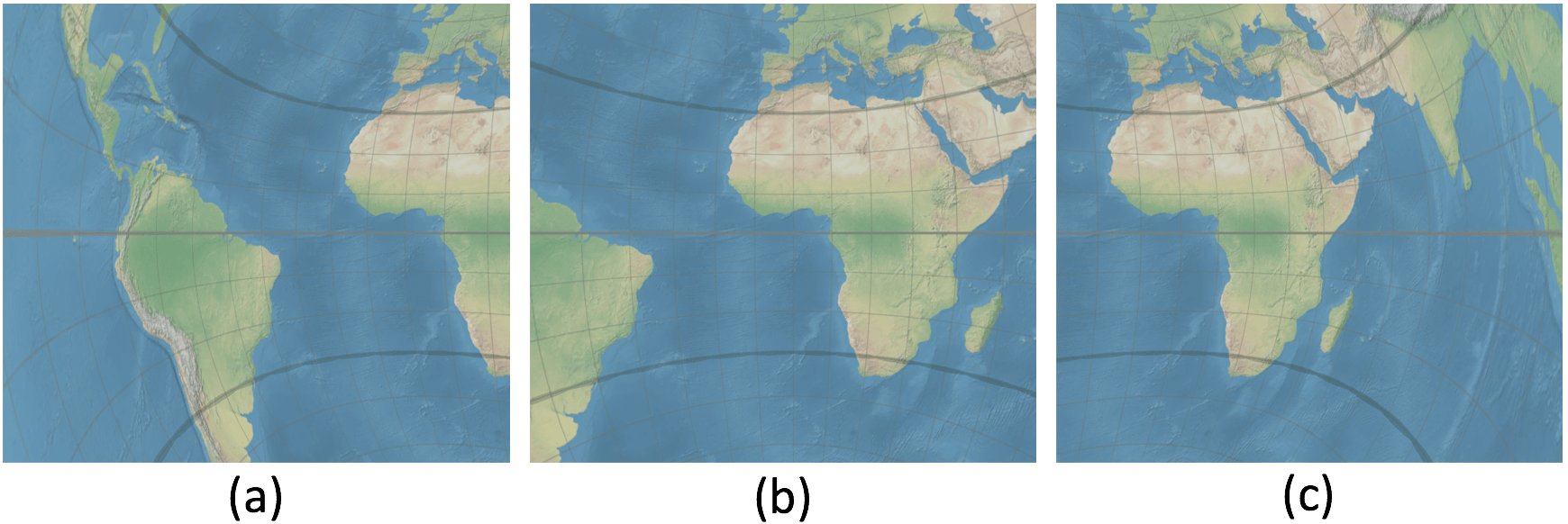}
	\vspace{-0.8em}
	\caption{Distortion of an egocentric globe changes with user position: (b) view at the initial location; (a) view from left close to the sphere's hull; (c) close view from right. %
	}
	\label{fig:lg-distortion}
	\vspace{-2.5em}
\end{figure}

The motivation for this design is to create a visualisation for VR with a maximum immersive experience. Areas, distances and shapes are not distorted on the egocentric sphere model, as there is no map projection involved.  When the head is positioned at the center of the sphere, there is little perceptual distortion, but if the user moves closer to the sphere and views it under an oblique angle, there is considerable distortion (see Fig.~\ref{fig:lg-distortion}). Slightly more than half of the sphere is visible to the user in the initial position.

In our pilot study, participants reported experiencing motion sickness in the egocentric globe. A stable external horizon reference is known to reduce motion sickness \cite{Bos:2006vf}. Thus, we added two static rings (Fig.~\ref{fig:lg-distortion}) to create a stable horizon. These rings align with two lines of constant latitude when the north pole is placed above the user. The rings remain in this static position when the user rotates the sphere. After adding this artificial horizon, participants reported that  dizziness was reduced and manageable.

\noindent\textbf{\textit{Curved Map:}}
We project an area-preserving map onto a spherical section (Fig.~\ref{fig:curvedmapteaser}). The user stands at the centre of this sphere, which has a radius of 1 metre. The map covers a horizontal angle of 108$^\circ$ and a vertical angle of 54$^\circ$. Using a section of a sphere as a projective surface is not a new idea in VR, e.g., Kwon et al. \cite{Kwon:2016go} use spherical projection surfaces for graph analysis. However, these previous applications did not use geospatial data or geographic maps.

The motivation for the curved map is to (a) create a more immersive experience than with flat maps, (b) allow the user to view the entire map with minimum head or eye movements, and (c) reduce perceptual distortion created by an oblique viewing angle. Our curved map is rendered with an additional ramp shader with a gradient texture to create the impression of a concave surface. 
As for the flat map, we use the Hammer projection when transforming the sphere to an initial flat map. We then apply a second transformation, which linearly maps the Hammer map onto the sphere. The resulting curved map does not preserve the relative size of areas.
 
\noindent\textbf{\textit{Interaction:}}
\label{sec:interaction}
We provide similar interaction across the four visualisations.
First, the VR model is fixed in space allowing the viewer to approach and move around it. 
Second, we allowed the viewer to adjust the centre of the geographic area in the visualisation. They could pick any location and drag it to a new position using a standard VR controller. The geographic location smoothly follows the beam that is sent from the controller and intersects the map or globe surface. A spherical rotation is applied to the geometry model before the model is projected (for the flat and curved maps) or rendered (for the exocentric and egocentric globes) \cite{Bostock:2013ix,Davies:2013ug}. With this interactive adjustment of the map centre, users can fine-tune the visualisations such that the area of interest is displayed at the centre of the map or globe.\footnote{A video demonstration of this interaction is included in this submission}
Adjusting the geographic centre allows for (1) bringing features of interest to the centre of the visualisation, (2) reducing distortion in shape of features of interest, and (3) avoiding the path or area of interest being split into two by the edge of the maps or out of view with the globes.

\added{A degree of zooming interaction is possible by the viewer physically moving closer to the surface of the map. For simplicity, the experiment was designed such that more extreme zooming was unnecessary. That is, all targets were sufficiently visible at the natural view position.  While other zoom interactions were not included in this study, it is worth noting that they would be required to support tasks where multiple view scales were necessary, e.g.\ moving from country to city scale.}

\vspace{-1.2em}
\section{User Study}
\label{sec:study}
Following~\cite{Battersby:2009jo,Friedman:2000ux,Friedman:2006ux,Carbon:2010bo} we evaluated four visualisations with three tasks essential to geospatial data visualisation: distance comparison, area comparison, and direction estimation.  \added{These tasks relate to real-world analysis scenarios, for example: area comparison is used to analyse the global forest cover change \cite{Hansen:2013iy}, global air quality monitoring \cite{Chu:2003ca}; distance comparison and direction estimation are used to help analyse the global movement of ocean animals \cite{Block:2011be}, cargo ship movements\cite{Kaluza:2010hb}, and air transportation networks \cite{Guimera:2005dm}.}
\vspace{-0.1em}
\noindent\textbf{Stimuli and tasks}\\
In order to rule out the influence of previous geographic knowledge, we avoided basing tasks on real-world geographic features, such as comparing the size of two existing countries, and used artificial distances and areas instead.

\vspace{-0.1em}
\noindent\textbf{\textit{Distance Comparison Task}:} Given two pairs of points, find the pair separated by the largest (spherical great circle) distance. 
Following Feiner \emph{et al.}~\cite{Feiner:1993ip}, we use leader lines to link the points with labels: ``A'', ``B'' for one pair and ``X'', ``Y''  for the other. 
Labels were kept horizontal and oriented towards the viewer via rotation in real time. 
Two factors that may affect user performance:
\emph{Variation} of the relative distance between each pair of points;
\emph{Geographic distance} between geographic midpoints of the two pairs.

The coefficient of variation (CV)~\cite{Brown:1998dj} was used to measure the variation between the distances of two pairs. We designed three groups of tasks with different difficulty levels:
\emph{easy}, with large variation in the pairwise distances and close geographic distance between the pairs;
\emph{small variation} between the pairwise distances and close geographic distance between the pairs;
\emph{far distance} with large variation in the pairwise distance and far geographic distance between the pairs.
In our pilot testing, far distance with small variation was found to be too difficult.

Initially, 20\% and 10\% CV were used as large and small variations, however, the resulting error rates were very low in our first pilot study. We therefore adjusted the CV values to 10\% and 5\%, respectively.
For the geographic distance, we chose 60$^\circ$ for the short distance, which approximately spans the size of a continent, and 120$^\circ$ was chosen for the long distance.
Two pairs of points were randomly generated. The distance between a pair of points is restricted to lie  in the range of $40^\circ$-$~60^\circ$.

\vspace{-0.2em}
\noindent\textbf{\textit{Area Comparison Task}:} Given two labeled polygons, identify the polygon with the larger area. Steradian (spherical area) was used to calculate the reference areas of polygons. The  leader line system was again used to link the centroids of the polygons with labels.
As with the distance comparison task, user performance may be affected by: \emph{variation} between the the areas of the polygons; \emph{geographic distance} between the centroids of polygons.

Again, three groups of tasks of different difficulty levels were designed. 20\% and 10\% CV were initially used as large and small variations, but the error rates were very low in the first pilot study. We adjusted the CV values to 10\% and 5\%, respectively, but participants found small variation tasks to be too difficult. Finally, we changed the CV values to 10\% and 7.5\%, respectively. As for the distance comparison tasks, 60$^\circ$ and 120$^\circ$ were used to distinguish the geographic distance between the centroids of the two polygons.

A convex polygon was created by linking eight random points generated with the same geographic distance (8$^\circ$) from a centroid. To ensure the generated polygons are in similar shape, the minimum central angle between two adjacent points was 30$^\circ$. To avoid the effect of color on area perception, the two polygons were placed on a similar background (either both in the sea or both on land).

\vspace{-0.2em}
\noindent\textbf{\textit{Direction Estimation Task}:} Given a short arrow, estimate whether or not the path continuation passes through a given location. 
The main factor that might affect user performance is
\emph{geographic distance} between the two locations.
Again, 60$^\circ$ and 120$^\circ$ were used to distinguish the geographic distance. Two groups with different difficulty levels were designed for direction estimation tasks.

\begin{figure}
	\centering
	\includegraphics[width=0.4\textwidth]{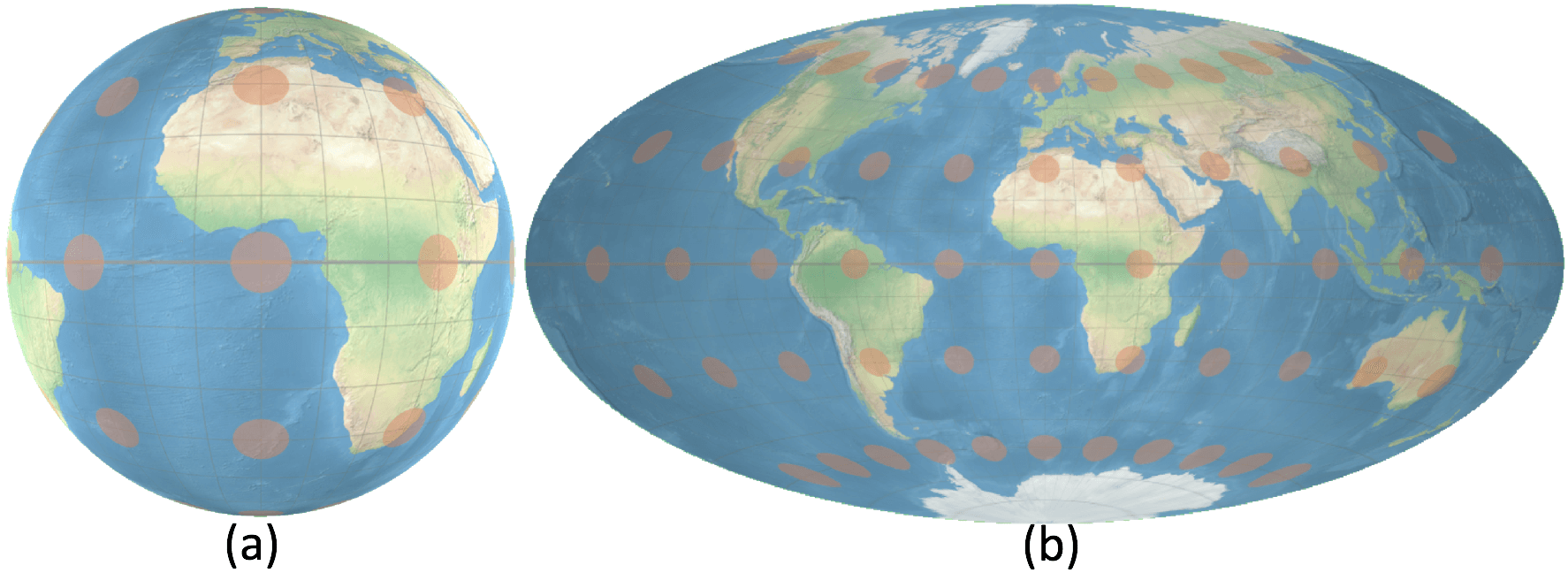}
	\vspace{-0.8em}
	\caption{Tissot indicatrices on the Natural Earth raster map with a graticule, (a) on an exocentric globe, (b) on a flat map.}
	\label{fig:tissot}
	\vspace{-2.5em}
\end{figure}

\vspace{-0.2em}
\noindent\textbf{Measures: response time, accuracy and interactions}\\
The response time was the interval between the initial rendering of the visualisation and the double-click of the trigger button on the controller. After double-click, the visualisation was hidden
and two answer options shown. As our tasks provide binary options we used the accuracy score from \cite{Willett:2015fv}: $(\frac{number\ of\ correct\ responses}{number\ of\ total\ responses} - 0.5) / 0.5$, where 1 indicates a perfect performance and 0 indicates a result equal to chance (i.e. randomly guessing).

We also measured the degree of interaction. The number of clicks has been widely used in conventional interaction evaluations, as well as in VR \cite{Kwon:2016go}. However, unlike those systems, the HTC Vive allows users to move in a larger open space, thus, not only the  number of interactions, but also the user's movements need to be considered. In our experiment, the positions and rotations of the user's head and the controller were recorded every 0.1s while the participants were viewing the visualisations. We used the aggregated changes in positions (in euclidean distance) and rotations (in degrees) between records to analyse user interactions.

\noindent\textbf{Experimental Setup}\\
An HTC Vive with 110$^\circ$ field of view and 90Hz refresh rate \cite{Dempsey:2016hn} was used as the VR headset in the experiment. The PC was equipped with an NVIDIA GeForce GTX 1080 graphics card and Intel i7-6700K 4.0GHz processor. Only one controller was needed in the experiment; a pointer from the controller (a shooting beam) was available at all times. We configured all interactions to work with the controller trigger. Participants could hold the trigger and manipulate the geographic centre.  The framerate was around 110FPS throughout the experiment, i.e.\ computation was faster than the display refresh rate. 
The Natural Earth raster map from \url{naturalearthdata.com} was used as the base texture for all tasks. A graticule was created for every 10$^\circ$ both in longitude and latitude. A thicker line was used for the equator. Tissot indicatrices (discussed below) shown in the training were rendered every 30$^\circ$ longitude and latitude (Fig.~\ref{fig:tissot}). The initial position of the globes and maps for each question was adjusted to ensure the user was looking at 0$^\circ$ longitude and 0$^\circ$ latitude.

\noindent\textbf{Participants}\\
We recruited 32 participants (11 female) from our university campus, all with normal or corrected-to-normal vision. Participants included university students and researchers. 
22 participants were within the age group 20--30, 7 participants were between 30--40, and 3 participants were over 40. VR experience varied:  25 participants had less than 5 hours of prior VR experience, 6 participants had 6--20 hours, and 1 participant had more than 20 hours. 

\noindent\textbf{Design}\\
The experiment was within-subjects: 32 participants $\times$ 4 visualisations $\times$ 3 tasks $\times$ 9 repetitions = 3,456 questions (108 questions per participant) with performance measures and lasted one hour on average. Each of the three difficulty conditions for distance and area comparison tasks was repeated 3 times. For direction estimation, we trialled 5 repetitions for the two difficulty conditions in the pilot study. Pilot participants reported the direction estimation for long distance was too difficult. We modified the repetition of the direction estimation to 3 repetitions for far distance and 6 for close distance. 
The mapping between tasks and techniques was counterbalanced across subjects by keeping the order of tasks constant (in the order of distance comparison, area comparison, and direction estimation) and using a Latin square design to balance the order of visualisations.

\noindent\textbf{Procedure}\\
Participants were first given a brief introduction about this project, the four types of visualisations, and the three types of tasks. Two types of training were included in this experiment: interaction training and task training.

\emph{Interaction training} was conducted when each visualisation was presented to the participants for the first time. The participants were introduced to the interactive visualisations and given sufficient time to familiarise themselves with the interaction. They were then asked to use the interaction to move Melbourne to the centre of their view and double-click on it. This activity familarised participants with the VR headset and controller, as well as each interactive visualisation.

\emph{Task training} was conducted when each condition (task $\times$ visualisation) was presented to the participants for the first time. Two sample tasks, different from the experimental tasks, were given to participants with unlimited time. After participants finished a training task, we highlighted the correct answer and presented additional geographic information related to the task when applicable: for distance comparison tasks, lines connecting each pair showing actual geographic distances were shown; for direction estimation tasks, actual geographic trajectories were shown. We reminded the participants to test their strategies both when they were doing the training tasks and when the correct answers were shown. For the two training direction estimation tasks, participants were presented with one ``hit''  and one ``miss'' condition. 

Participants were asked to finish the different types of tasks one by one. Within one task, they were presented the four visualisations one after the other. They were not explicitly informed about the area and angular distortion that can be caused by map projections, but during the training we displayed Tissot indicatrices (Fig.~\ref{fig:tissot}) to show scale variation and angular distortion on all four visualisations. A Tissot indicatrix appears as a circle on maps without angular distortion, and appears as an ellipse on maps with angular distortion. The size of Tissot indicatrices changes with area distortion, resulting in larger ellipses where area is inflated, and smaller ellipses where area is compressed \cite{Snyder:1987tk}. Tissot indicatrices were not shown during the experimental tasks.

\noindent A posthoc questionaire recorded feedback on:\\
- preference ranking of visualisations in terms of visual design and ease of use for the experimental tasks;\\
- experience of motion sickness in the different visualisations;\\
- advantages and disadvantages for each visualisation;\\
- strategies for different types of tasks;\\
- background information about the participant.

\vspace{-1.5em}
\section{Results}
Histograms and Q$-$Q plots revealed that the error rate distribution was not normal. As there were more than two conditions, we used the Friedman test to check for significance and applied the Wilcoxon-Nemenyi-McDonald-Thompson post-hoc test to conduct pairwise  comparisons \cite{Hollander:1999ns}. 
To compare response time and user interactions we considered only times and interactions for correct responses. Histograms and Q$-$Q plots showed both distributions were approximately normal. Due to the unbalanced number of correct answers per participant, we chose linear mixed-effects (LME) ANOVA to check for significance and applied Tukey's HSD post-hoc tests to conduct pairwise comparisons \cite{McCulloch:2013kl,Pinheiro:2000id}.
For user preferences and motion sickness ratings we again used the Friedman test and the Wilcoxon-Nemenyi-McDonald-Thompson post-hoc test to test for significance.

\begin{figure}
	\centering
	\includegraphics[height=3.5cm]{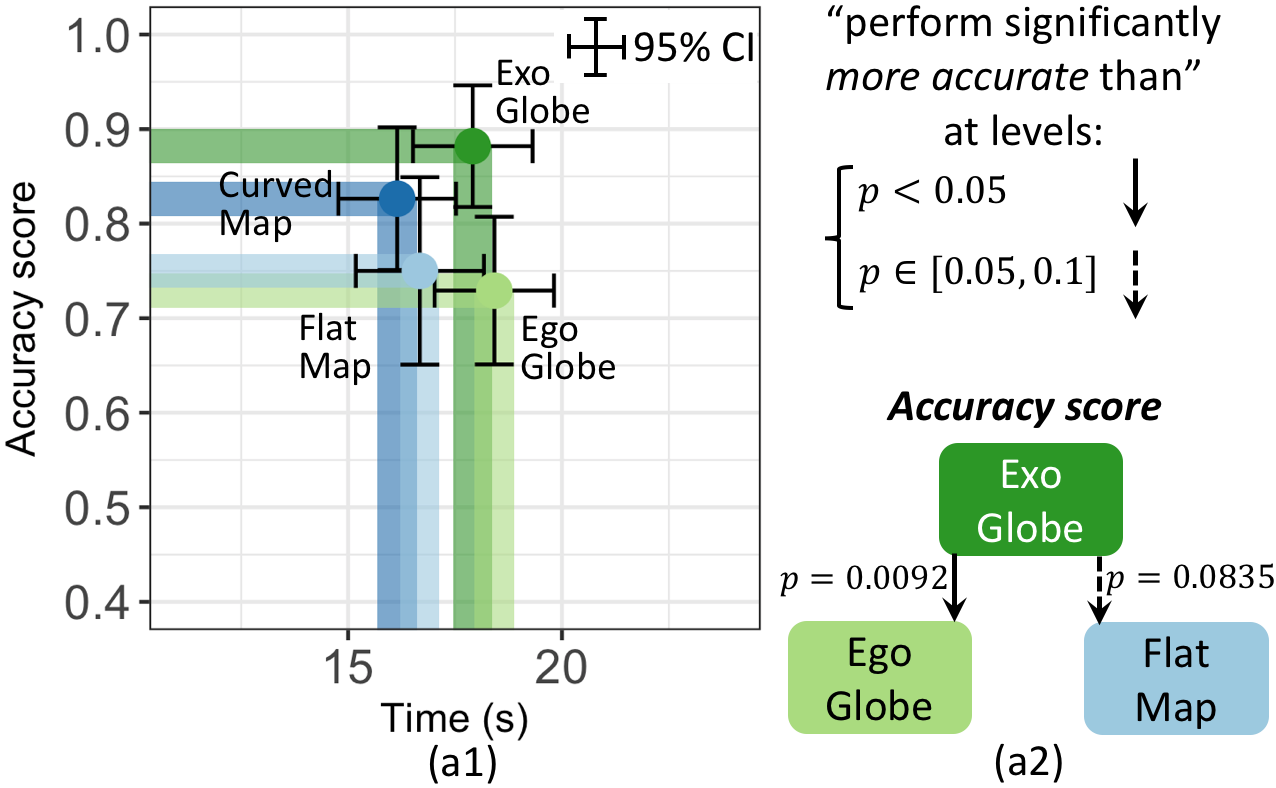}
	\vspace{-0.8em}
	\caption{(a1) Average performance of \textbf{distance comparison} task per visualisation, with 95\% confidence interval, (a2) graphical depiction of results of pairwise post-hoc test.}
	\label{fig:distance-vis}
	\vspace{-2.5em}
\end{figure}

\begin{figure}[t]
	\centering
	\includegraphics[width=0.45\textwidth]{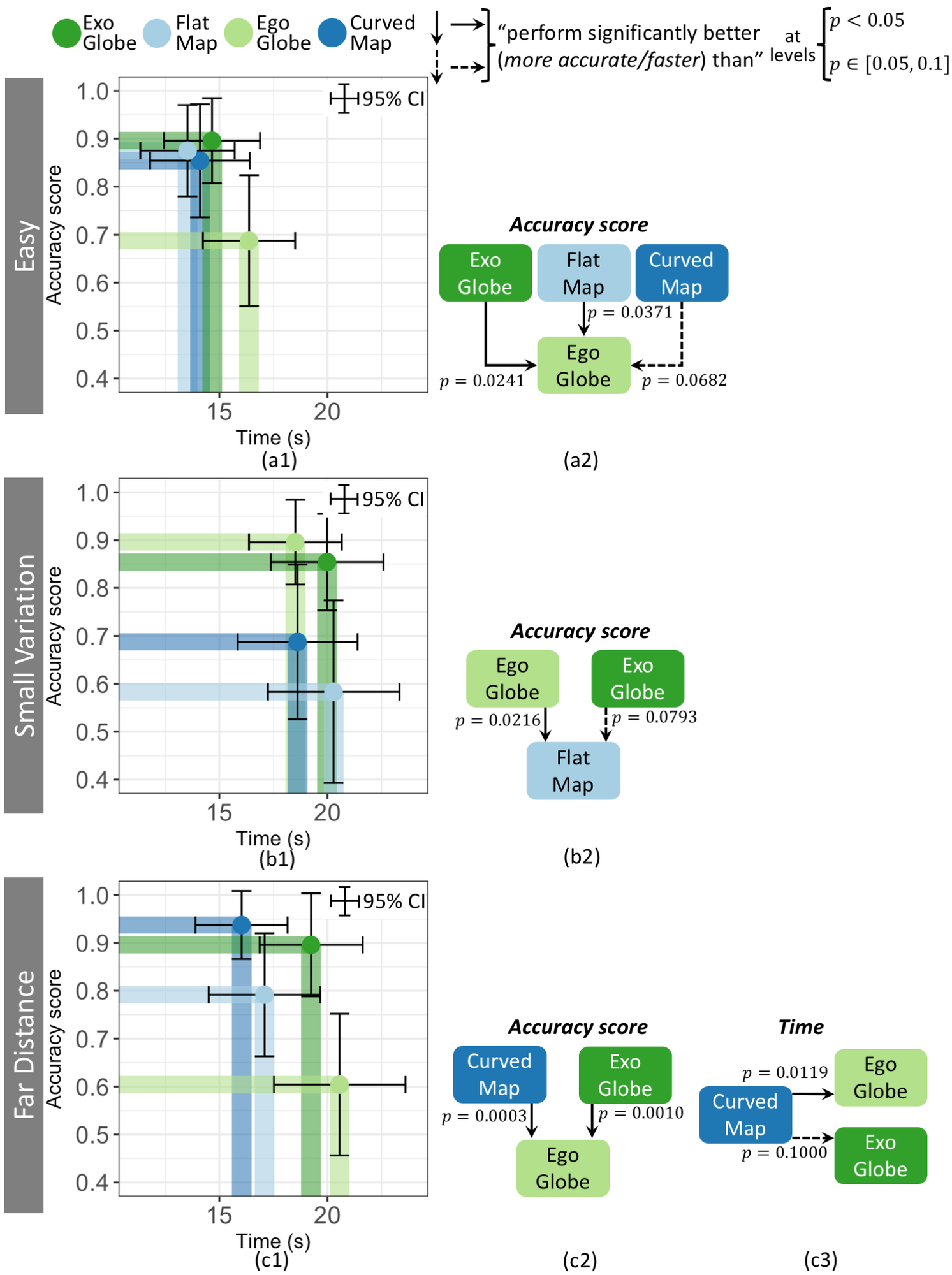}
	\vspace{-0.8em}
	\caption{Break down of \textbf{distance comparison} task into different difficulty conditions. (a1,b1,c1) Average performance per visualisation with 95\% confidence interval, (a2,b2,c2,c3) graphical depiction of result of pairwise post-hoc test.}
	\label{fig:distance-vis-break}
	\vspace{-2.8em}
\end{figure}

\noindent\textbf{Distance Comparison}\\
The Friedman test revealed a statistically significant effect of visualisations on accuracy ($\chi^2(3) = 11.453, p = .0095$). Fig.~\ref{fig:distance-vis}(a1) shows the average accuracy score of exocentric globe (0.88) was higher than that of egocentric globe (0.73) and flat map (0.75). While curved map (0.83) outperformed egocentric globe and flat map, this was not found to be statistically significant. A post-hoc test showed statistical significances as per Fig.~\ref{fig:distance-vis}(a2). 
The LME ANOVA analysis showed significant effect of visualisations on time ($\chi^2(3) = 6.837, p = .0773$). Fig.~\ref{fig:distance-vis}(a1) shows avg.\ response times with curved map (16.1s) and flat map (16.7s) were less than the avg.\ response times for exocentric (17.9s) and egocentric globe (18.4s), however, the post-hoc test did not find statistical significance.  By difficulty condition the Friedman test revealed significant effect for \textbf{avg.\ accuracy}:\\
\noindent\textbf{\textit{Easy}:}
$\chi^2(3)=10.711, p = .0134$. Fig.~\ref{fig:distance-vis-break}(a1) shows avg.\ accuracy of egocentric globe (0.69) $<$ exocentric globe (0.90), flat map (0.88), and curved map (0.86). Significance in post-hoc testing detailed in Fig.~\ref{fig:distance-vis-break}(a2).\\
\noindent\textbf{\emph{Small variation}:} 
$\chi^2(3)=10.938, p = .0120$. Fig.~\ref{fig:distance-vis-break}(b1): avg.\ accuracy of flat map (0.58) $<$ egocentric globe (0.90) and exocentric globe (0.86), post-hoc significance Fig.~\ref{fig:distance-vis-break}(b2)). Curved map accuracy (0.69) is similar to flat map.\\
\noindent\textbf{\textit{Far distance}:} 
$\chi^2(3)=20.131, p = .0001$. Fig.~\ref{fig:distance-vis-break}(c1): avg.\ accuracy of egocentric globe (0.60) $<$ curved map (0.94) and exocentric globe (0.90). Post-hoc testing indicated statistical significance (see Fig.~\ref{fig:distance-vis-break}(c2)). Avg.\ accuracy of flat map (0.79) is higher than that of the egocentric globe, however, the post-hoc test did not find statistical significance.

\noindent For \textbf{avg. response time}, LME ANOVA analysis revealed:\\
\noindent\textbf{\textit{Easy}:} 
no significance ($\chi^2(3) = 4.903, p = .179$). Although in Fig.~\ref{fig:distance-vis-break}(a1), egocentric globe (16.3s) tended to be slower than others (flat map 13.5s, exocentric globe 14.7s and curved map 14.1s). \\
\noindent\textbf{\textit{Small variation}:} 
no significance ($\chi^2(3) = 1.147, p = .765$).\\
\noindent\textbf{\textit{Far distance}:}
a significant effect ($\chi^2(3) = 10.451, p = .0151$). Fig.~\ref{fig:distance-vis-break}(c1) shows curved map (16.0s) was faster than either exocentric globe (19.2s) or egocentric globe (20.6s). Post-hoc: Fig.~\ref{fig:distance-vis-break}(c2)). Flat map response time (17.1s) was similar to curved map.

\noindent\textbf{\textit{Participant strategies}:} From the questionnaires, we found two general strategies for distance comparison with all visualisations:\\
\noindent\textit{ - Using the graticule grid to calculate distance} -- two explicitly mentioned using Manhattan distance;\\
\noindent\textit{ - Moving each pair in turn to the centre of the map or globe} -- usually  more than once and alternating between them.

\noindent There were also specific strategies for different visualisations:\\
\noindent\textit{ - For flat and curved maps:} moving the two pairs so they are placed symmetrically around the centre. ``It seems they will have the same scale''. This strategy was not usually possible in the globes because of the more limited field of view.\\
\noindent\textit{ - For exocentric and egocentric globes:} some participants used proprioception to estimate the distance between the points in each pair. They used interaction to rotate the sphere to move the points past a reference point (such as the center of the globe) and the  effort required for each pair gave an estimate of the relative distance. \\
\noindent\textit{ - For the exocentric globe:} use the top of the visible hemisphere as a reference point, in turn placing one point of each pair at this position and memorising the position of the other point in the pair, then switching to the other pair. Often they ensured  the pairs were vertically aligned.

\begin{figure}
	\centering
	\includegraphics[height=3.5cm]{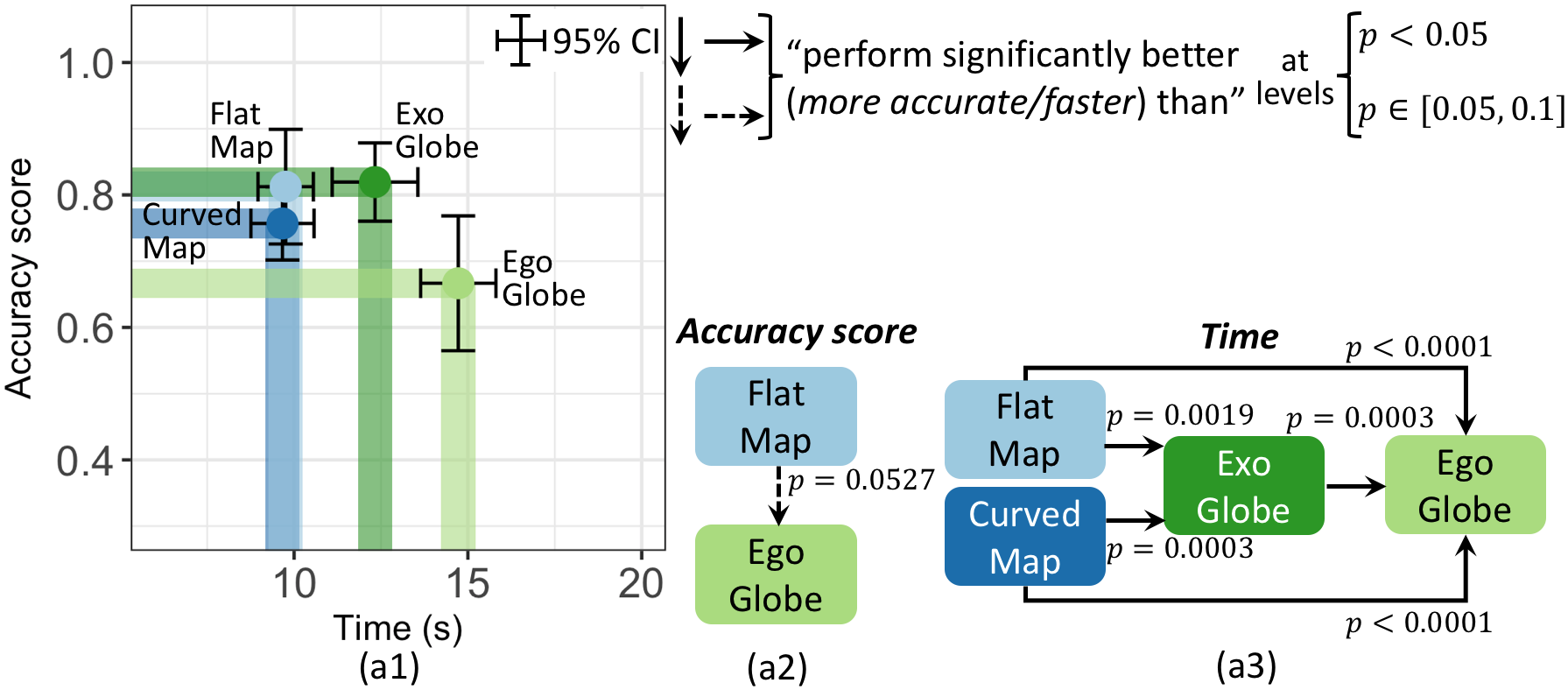}
	\vspace{-0.8em}
	\caption{(a1) Average performance of \textbf{area comparison} task per visualisation, with 95\% confidence interval, (a2, a3) graphical depiction of results of pairwise post-hoc test.}
	\label{fig:area-vis}
	\vspace{-2.3em}
\end{figure}

\begin{figure}
	\centering
	\includegraphics[width=0.45\textwidth]{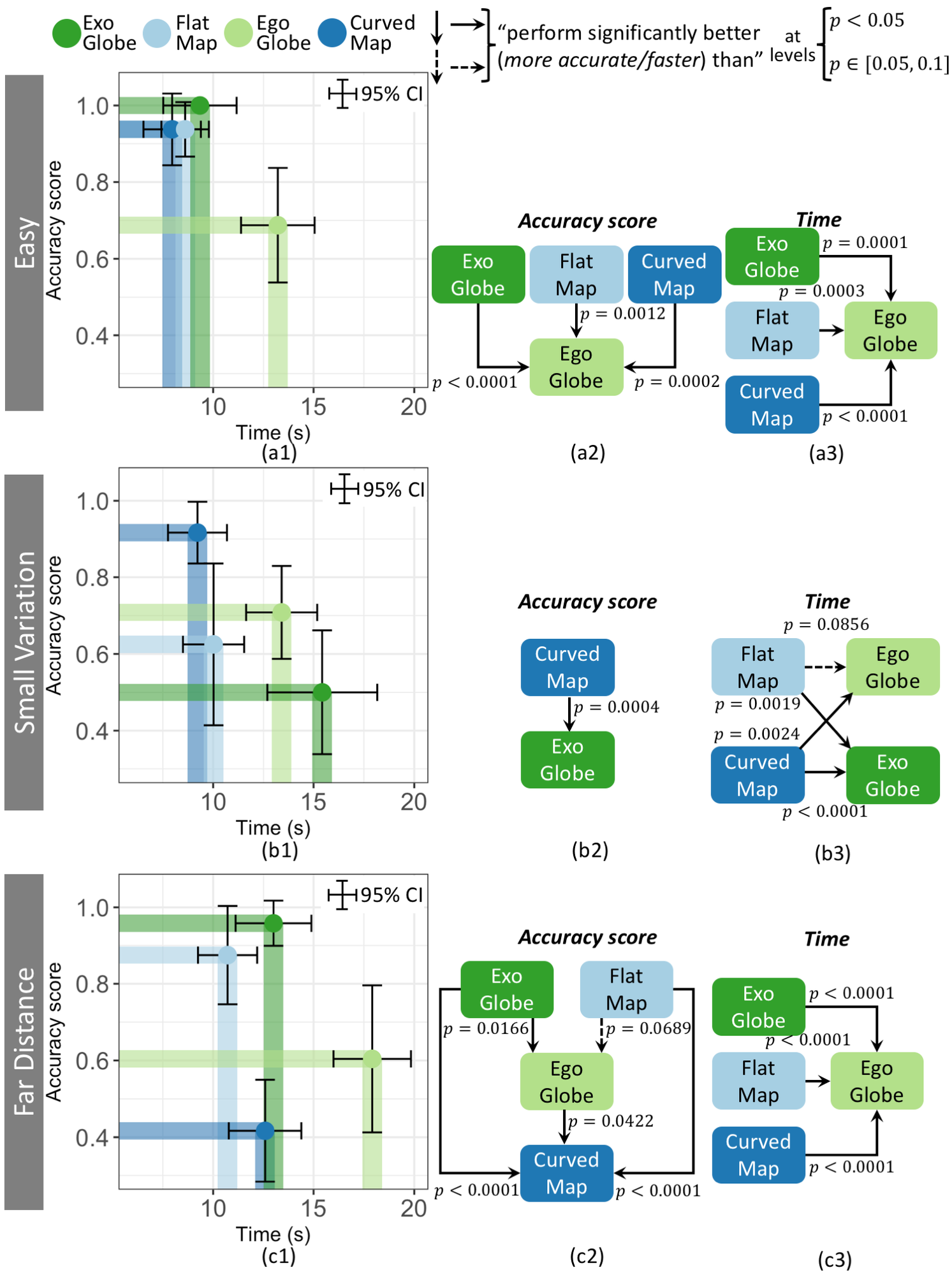}
	\vspace{-0.8em}
	\caption{Break down of \textbf{area comparison} task into different difficulty conditions. (a1,b1,c1) Average performance per visualisation with 95\% confidence interval, (a2,a3,b2,b3,c2,c3) graphical depiction of results of pairwise post-hoc test.}
	\label{fig:area-vis-break}
	\vspace{-2.5em}
\end{figure}

\noindent\textbf{Area Comparison}\\
The Friedman test revealed a statistically significant effect of visualisations on accuracy for the area comparison task ($\chi^2(3)=7.218, p = .0652$). Fig.~\ref{fig:area-vis}(a1) shows the average accuracy score of egocentric globe (0.67) was lower than that of the others (exocentric globe 0.82, flat map 0.81, and curved map 0.76). The post-hoc test only shows a significant difference between the comparison of flat map and egocentric globe (see Fig.~\ref{fig:area-vis}(a2)).
The LME ANOVA analysis revealed a statistically significant effect of visualisations on time ($\chi^2(3)=46.762, p < .0001$). Fig.~\ref{fig:area-vis}(a1) shows the average response times with curved map (9.6s) and flat map (9.7s) were less than those of exocentric globe (12.3s) and egocentric globe (14.7s). The exocentric globe also performed significantly faster than egocentric globe. A post-hoc test showed statistical significances as per Fig.~\ref{fig:area-vis}(a3). By difficulty condition the Friedman test revealed significant effect for \textbf{avg. accuracy}:\\
\noindent\textbf{\textit{Easy}:} 
$\chi^2(3)=27.545, p < .0001$. Fig.~\ref{fig:area-vis-break}(a1) shows the avg.\ accuracy score of egocentric globe (0.69) < the other visualisations (exocentric globe 1.00, flat map 0.94, and curved map 0.94). A post-hoc test showed statistical significances as per Fig.~\ref{fig:area-vis-break}(a2). \\
\noindent\textbf{\textit{Small variation}:} 
$\chi^2(3)=15.451, p = .0014$. Fig.~\ref{fig:area-vis-break}(b1) shows the avg.\ accuracy score of curved map (0.92) > the other visualisations (egocentric globe 0.71, flat map 0.63, and excocentric globe 0.50). The post-hoc test showed statistical significance for the difference between curved map and exocentric globe (see Fig.~\ref{fig:area-vis-break}(b2)). \\
\noindent\textbf{\textit{Far distance}:} 
$\chi^2(3)=39.346, p < .0001$. Fig.~\ref{fig:area-vis-break}(c1) shows the avg.\ accuracy scores with excocentric globe (0.96) and flat map (0.88) > egocentric globe (0.60) and curved map (0.42). The egocentric globe also performed more accurately than curved map. A post-hoc test showed statistical significances as per Fig.~\ref{fig:area-vis-break}(c2).

\noindent LME ANOVA revealed significant effect for \textbf{avg. response time}:\\
\noindent\textbf{\textit{Easy}:}
$\chi^2(3)=31.269, p < .0001$. Fig.~\ref{fig:area-vis-break}(a1) shows egocentric globe (13.2s) was slower than that of the other visualisations (curved map 7.9s, flat map 8.6s, and exocentric globe 9.3s). A post-hoc test showed statistical significances as per Fig.~\ref{fig:area-vis-break}(a3). \\
\noindent\textbf{\textit{Small variation}:}
$\chi^2(3)=21.663, p < .0001$. Fig.~\ref{fig:area-vis-break}(b1) shows curved map (9.2s) and flat map (10.0s) were faster than those of egocentric globe (13.4s) and exocentric globe (15.4s). A post-hoc test showed statistical significances as per Fig.~\ref{fig:area-vis-break}(b3). \\
\noindent\textbf{\textit{Far distance}:}
$\chi^2(3)=36.511, p < .0001$. Fig.~\ref{fig:area-vis-break}(c1) shows egocentric globe (17.91s) was slower than those of the other visualisations (flat map 10.7s, curved map 12.5s, and  exocentric globe 17.9s). A post-hoc test showed statistical significances as per Fig.~\ref{fig:area-vis-break}(c3)).

\noindent\textbf{\textit{Participant strategies}:} 
From the questionnaire we found two general strategies for area comparison used for all visualisations \\
\noindent\textit{ - Using the graticule grid} to estimate the area. \\
\noindent\textit{ -  Moving each polygon in turn} to the centre of the map or globe, usually  more than once and alternating between them.

\noindent One specific strategy for flat and curved maps was identified:\\
\noindent\textit{ - For flat and curved maps}, moving the two polygons so they are placed symmetrically around the centre. This strategy was not usually possible in the exocentric globes and egocentric globes because of the more limited field of view.

\begin{figure}
	\centering
	\includegraphics[height=3.5cm]{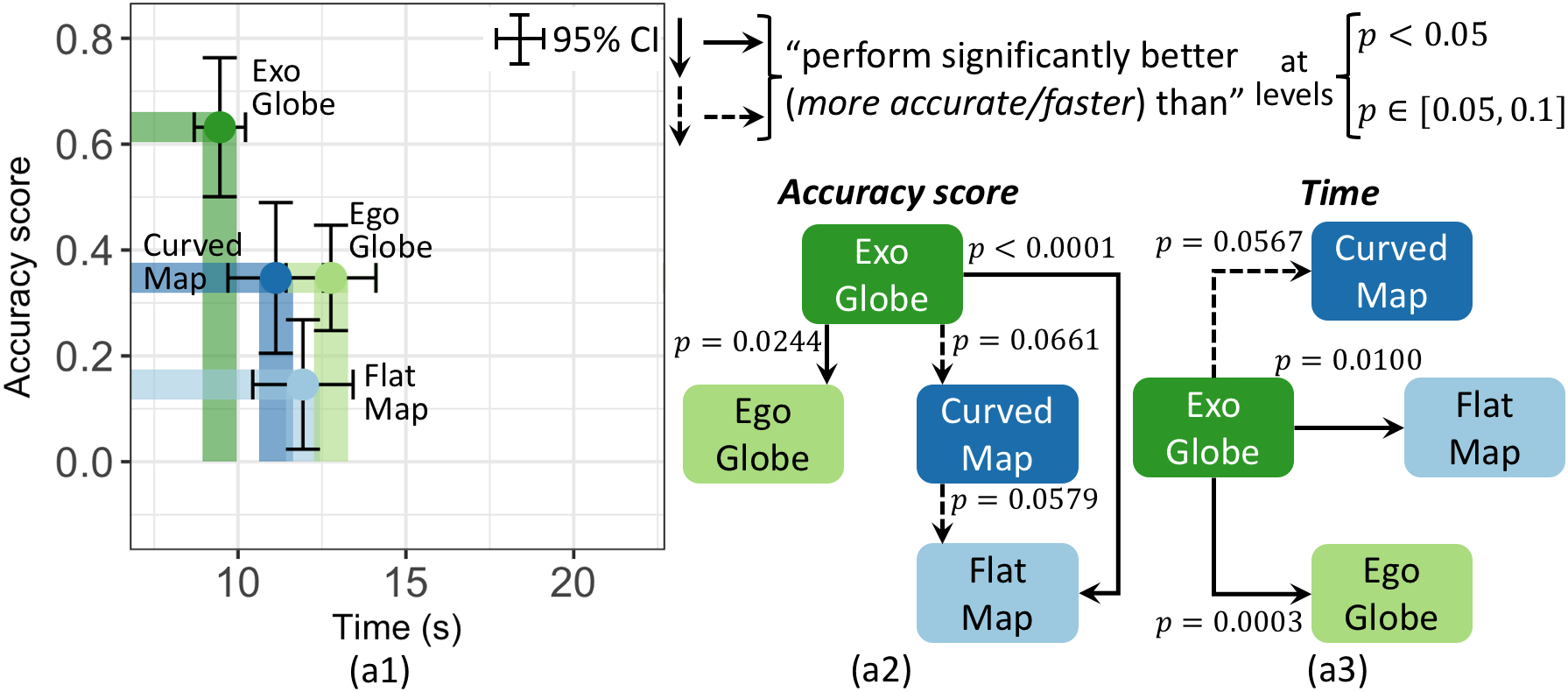}
	\vspace{-0.8em}
	\caption{(a1) Average performance of \textbf{direction estimation} task per visualisation, with 95\% confidence interval, (a2, a3) graphical depiction of results of pairwise post-hoc test.}
	\label{fig:nav-vis}
	\vspace{-1.5em}
\end{figure}

\begin{figure}
	\centering
	\includegraphics[width=0.45\textwidth]{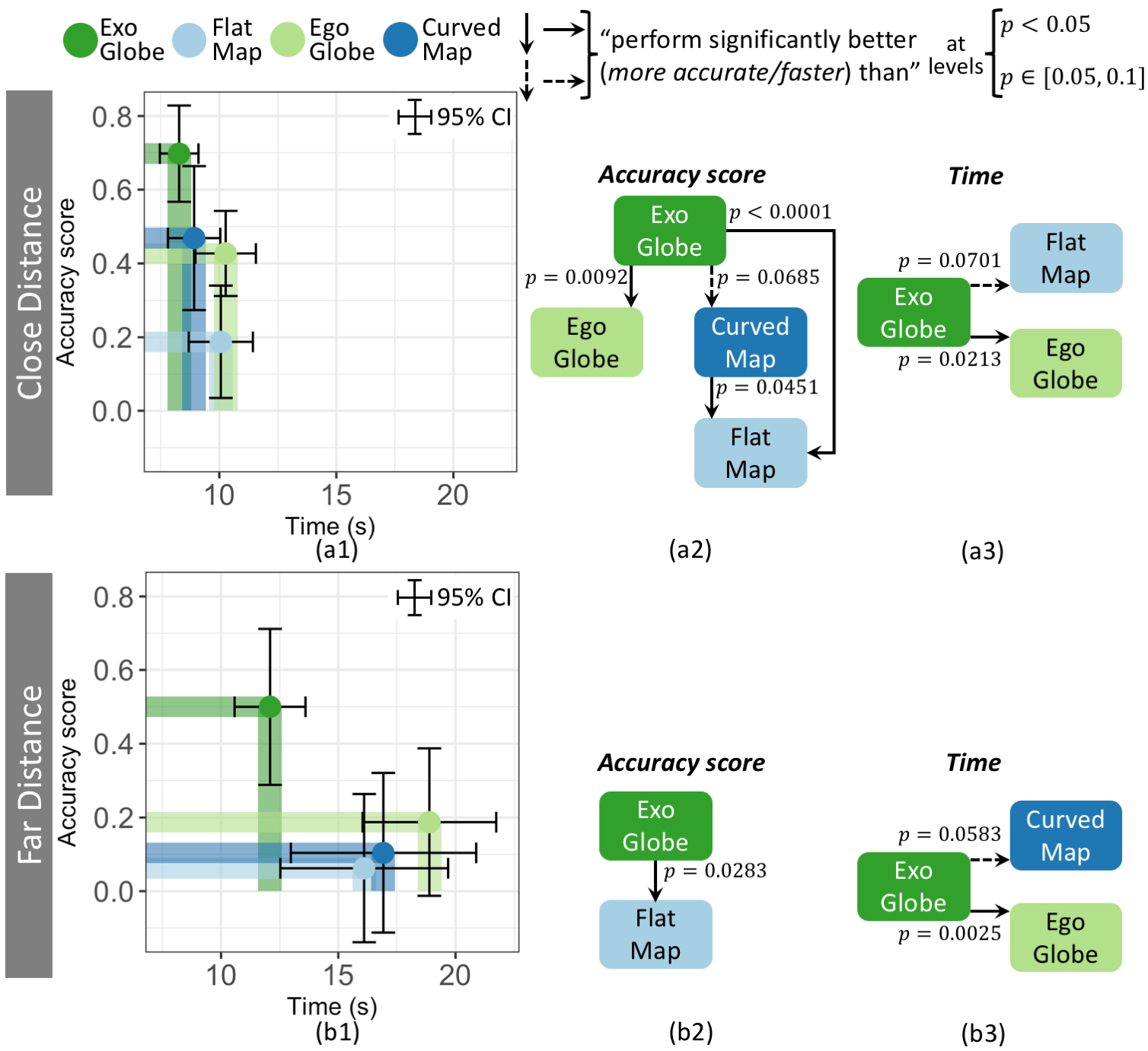}
	\vspace{-0.8em}
	\caption{Break down of \textbf{direction estimation} task into different difficulty conditions. (a1,b1) Average performance per visualisation with 95\% confidence interval, (a2,a3,b2,b3) graphical depiction of results of pairwise post-hoc test.}
	\label{fig:nav-vis-break}
	\vspace{-3em}
\end{figure}

\noindent\textbf{Direction Estimation}\\
The Friedman test revealed a statistically significant effect of visualisations on accuracy for the direction estimation task ($\chi^2(3)=24.937, p < .0001$). Fig.~\ref{fig:nav-vis}(a1) shows the average accuracy score of exocentric globe (0.63) is higher than those of the other visualisations (curved map 0.35, egocentric globe 0.35, and flat map 0.15). The post-hoc test showed statistical significance for these differences. The curved map also had a statistically significantly higher accuracy score than the flat map (see Fig.~\ref{fig:nav-vis}(a2)).
The LME ANOVA analysis revealed a statistically significant effect of visualisations on time ($\chi^2(3)=11.846, p = .0079$). In Fig.~\ref{fig:nav-vis}(a1), we can see the avg.\ response time with exocentric globe (9.4s) was faster than curved map (11.1s), flat map (11.9s), and egocentric globe (12.7s). The post-hoc test showed statistical significance (see Fig.~\ref{fig:nav-vis}(a3)). By difficulty condition the Friedman test revealed significant effect for \textbf{avg. accuracy}:\\
\noindent\textbf{\textit{Close distance}:} 
$\chi^2(3)=26.162, p < .0001$. Fig.~\ref{fig:nav-vis-break}(a1) shows the avg.\ accuracy score of exocentric globe (0.70) > other visualisations (curved map 0.47, egocentric globe 0.43, and flat map 0.19). Curved map also had a significantly higher accuracy score than flat map. A post-hoc test showed statistical significances as per Fig.~\ref{fig:nav-vis-break}(a2). \\ 
\noindent\textbf{\textit{Far distance}:}
$\chi^2(3)=8.496, p = .0368$. Fig.~\ref{fig:nav-vis-break}(b1) shows the avg.\ accuracy score of the exocentric globe (0.50) > other visualisations (egocentric globe 0.19, curved map 0.10, and flat map 0.06). The post-hoc test only showed statistical significance between the exocentric globe and the flat map (see Fig.~\ref{fig:nav-vis-break}(b2)).

\noindent LME ANOVA revealed significant effect for \textbf{avg. response time}:\\
\noindent\textbf{\textit{Close distance}:}
$\chi^2(3)=7.444, p = .0590$. Fig.~\ref{fig:nav-vis-break}(a1) shows texocentric globe (8.2s) was significantly faster than those of flat map (10.0s) and egocentric globe (10.2s). The exocentric globe tended to outperform the curved map (8.9s), however, no statistical significance is evident from the post-hoc test. A post-hoc test showed statistical significances as per  Fig.~\ref{fig:nav-vis-break}(a3). \\
\noindent\textbf{\textit{Far distance}:}
$\chi^2(3)=10.335, p = .0159$. Fig.~\ref{fig:nav-vis-break}(b1) shows exocentric globe (12.0s) tended to be faster than flat map (16.1s), curved map (16.9s), and egocentric globe (18.8s). The post-hoc test showed the statistical significance between exocentric globe and curved map and egocentric globe (see Fig.~\ref{fig:nav-vis-break}(b3)), but no statistical significance with flat map.

\noindent\textbf{\textit{Participant strategies}:}
From the questionnaire we identified three general strategies for direction estimation used for all visualisations:\\
\noindent\textit{ - Mentally following the arrow} -- Those participants ``attempted to `sit behind' the direction of the arrow.'' and ``imagined a marble running down the arrow.''\\
\noindent\textit{ - Moving the mid-point of the two points} -- to the centre of the map or globe.\\
\noindent\textit{ - Moving the start point of the arrow to the centre} -- some participants also tried to vertically/horizontally align the arrow.

\noindent There was one specific strategy for the exocentric globe and the egocentric globe:\\
\noindent\textit{ - Placing the start point of the arrow at the center of the globe}, then rotating the globe following the direction of the arrow.
\begin{figure}
	\centering
	\includegraphics[height=3.3cm]{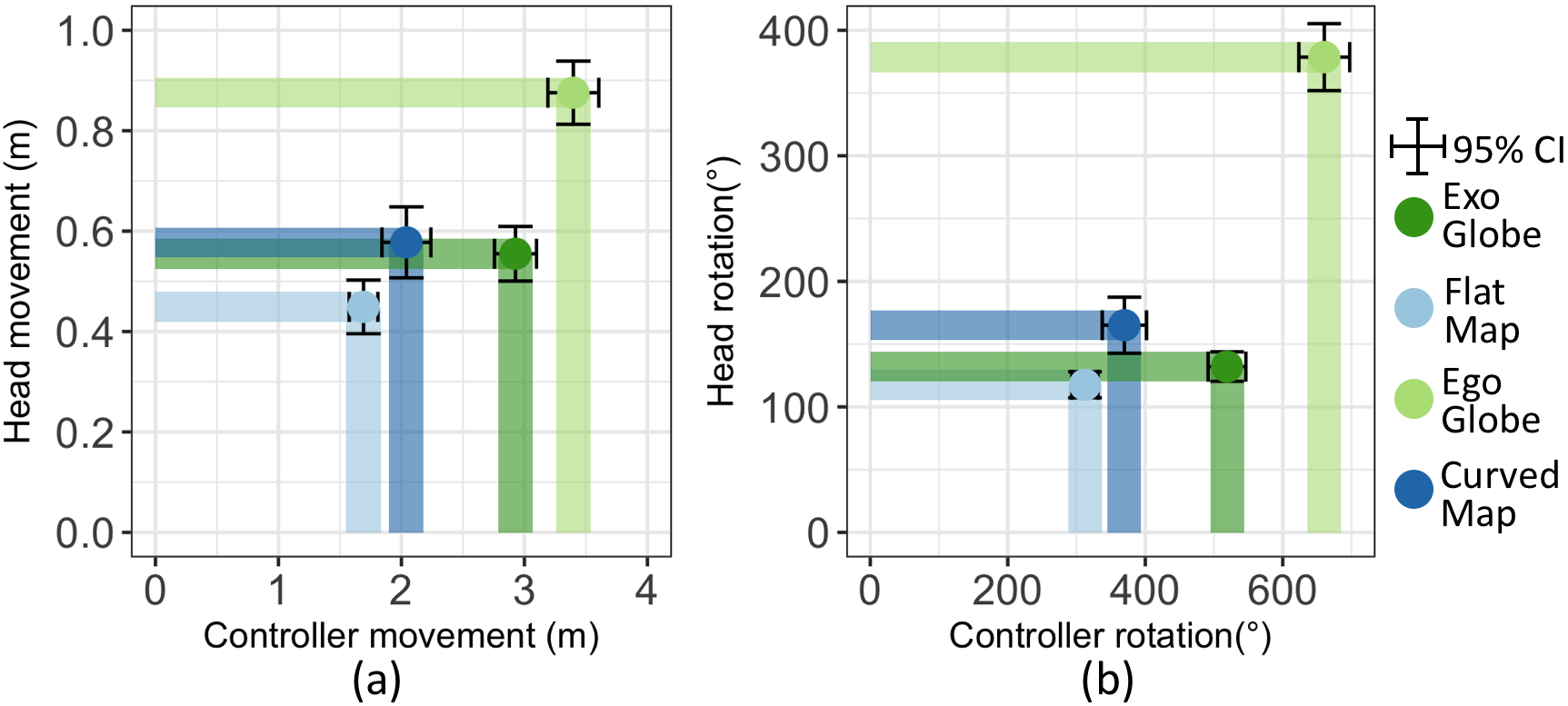}
	\vspace{-0.8em}
	\caption{(a) Average accumulated movements of head and controller per task; (b) average accumulated rotations of head and controller per task.}
	\label{fig:interactions-result}
	\vspace{-2.5em}
\end{figure}

\noindent\textbf{User interactions}\\
We analysed aggregate change in positions (euclidean distance) and rotation (in degrees) of the user's head and the controller,
Fig.~\ref{fig:interactions-result}, %
all results significant by LME ANOVA ($p<.0001$). 
Pairwise post-hoc testing revealed significant differences: egocentric globe $>$ exocentric globe $>$ curved map $>$ flat map; with all $p<.05$.
For head movement, post-hoc test revealed significance: egocentric globe $>$ exocentric globe $\approx$ curved map $>$ flat map; with all $p<.05$, except $\approx$ means no significant differences. For the rotations of the head, the post-hoc test revealed significant differences: egocentric globe $>$ curved map $>$ exocentric globe $\approx$ flat map; with all $p<.05$, except again $\approx$ means no significant differences. 

\noindent\textbf{Motion sickness}\\
\begin{minipage}{.63\columnwidth} 
A five-point-Likert scale was used for rating participant motion sickness, from \setlength{\fboxsep}{1pt}\colorbox{noDizzy}{\strut no motion sickness} to \setlength{\fboxsep}{1pt}\colorbox{dizzy}{strong motion sickness}. The Friedman test revealed a significant effect of visualisations on motion sickness rating ($\chi^2(3)=53.84, p < .0001$). From the figure, we can see that the percentage of 
\end{minipage}
\begin{minipage}{3.0cm}
\centering
\vspace{0mm}
\small
\fontfamily{phv}\selectfont
Motion sickness rating\\
\vspace{1mm}
\includegraphics[width=3.2cm]{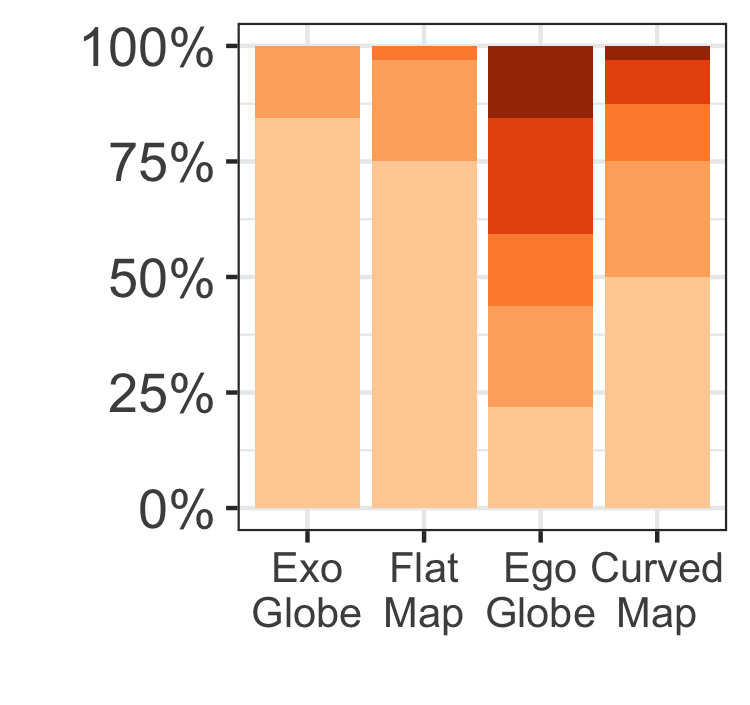}   
\vspace{-4mm}
\end{minipage}  
participants that did not experience
motion sickness in the egocentric globe (21.9\%) and the curved map (50\%) is significantly less than the percentage that did not experience motion sickness in the exocentric globe (84.4\%) and the flat map (75\%). The post-hoc test showed that participants experienced significantly more motion sickness in the egocentric globe and the curved map than they did in the exocentric globe and the flat map (all $p<.05$). The egocentric globe also caused significantly more motion sickness to participants than the curved map with $p=.0074$.

\noindent\textbf{User preference and feedback}\\
\noindent\begin{minipage}{0.63\columnwidth}
\vspace{0mm}
Participant ranking for each of the four visualisations by percentage of respondents is  shown by colour: \setlength{\fboxsep}{1.5pt}\colorbox{first}{\textcolor{white}{$1^{st}$}}, \setlength{\fboxsep}{1pt}\colorbox{second}{\textcolor{white}{$2^{nd}$}}, \setlength{\fboxsep}{1pt}\colorbox{third}{$3^{rd}$} and \setlength{\fboxsep}{1pt}\colorbox{fourth}{$4^{th}$}.
For \emph{visual design}, the Friedman test revealed a significant effect of visualisations on preference ($\chi^2(3)=10.612, p = .0140$). 
The strongest preference was for the exocentric globe, with 78.1\% voting it as first or second place. The preference 
\end{minipage}
\begin{minipage}{3.0cm}
\centering
\vspace{1mm}
\small
\fontfamily{phv}\selectfont
Visual Design Ranking\\
\vspace{0mm}
\includegraphics[width=3.2cm]{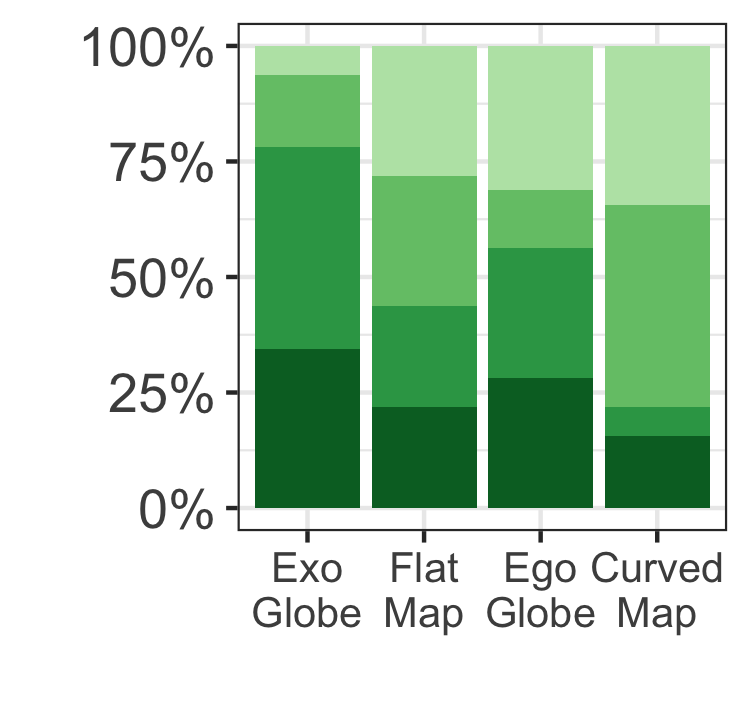} 
\vspace{-4mm}
\end{minipage} 
for the flat map and the egocentric globe were similar, each received 43.8\% and 56.3\% votes as first or second place. The curved map was the least preferred, with 21.9\% votes for first or second place. The post-hoc tests only showed a significant difference between the exocentric globe and the curved map with $p=.0080$.

\vspace{-2mm}
\noindent\begin{minipage}{0.63\columnwidth}
For \emph{readability}, the Friedman test indicated no significant effect of visualisations on preference ($\chi^2(3)=5.363, p = .1471$). The strongest preference is again for the exocentric globe, with 65.6\% of respondents voting it first or second place. The other visualisations have similar preferences, with the flat map, egocentric globe, 
\end{minipage}
\begin{minipage}{3.0cm}
\centering
\vspace{1mm}
\small
\fontfamily{phv}\selectfont 
Readability Ranking\\
\vspace{0mm}
\includegraphics[width=3.2cm]{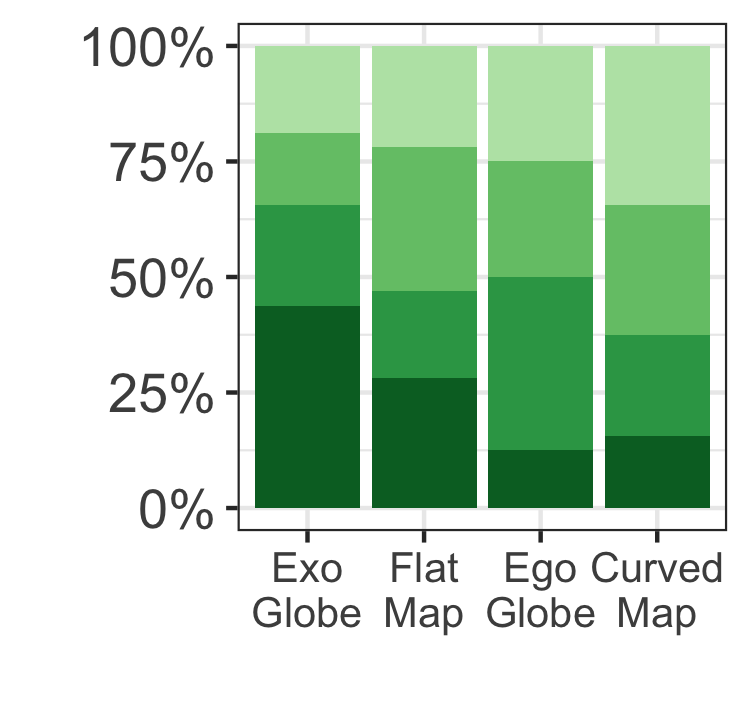}
\vspace{-4mm}
\end{minipage} 
and curved map receiving 46.9\%, 50\% and 37.5\% votes respectively as first or second place.

The final section of the study allowed participants to give feedback on the pros and cons of each design. Qualitative analysis of these comments reveal (overall):

\noindent\textbf{Exocentric globe} was intuitive, familiar and easy to manipulate: ``It is the same with a physical globe, the way to manipulate it is just nature, you can also walk around it.'', ``I feel like I am most habituated to this kind of representation''. Some also commented on the tasks: ``I think this one is the most suitable for solving the given tasks. It is not distorted at all.'' One noted, ``occlusion is a problem if targets are too far apart.'' \added{Note that even at 180 degrees separation, two points will still be visible at antipodes.  However, perceptual distortion due to curvature of the globe was problematic for our participants in the far condition.}
 Some also suggested a different interaction, which would synchronise the rotations of the exocentric globe with the hand rotations.

\noindent\textbf{Flat map} was again very familiar: ``Similar to what we see on paper''. Some enjoyed the ability to show the whole world at once. Others were excited about the interaction of manipulating the geographic centre: ``it works as fluid and has a feeling of artwork''. However, this interaction was unfamiliar to most participants: ``it took a while to get used to controlling the map in the way that I wanted to.'' Interacting with the map can result in landmasses with unusual orientation, which is also unfamiliar to most participants. Yet, some commented that it is boring to have a flat map in VR.

\noindent\textbf{Egocentric globe} was novel and immersive: ``It is cool and exciting being inside of the world''. However, ``I don't like it when it comes to answering questions''. Some also commented ``it is easier to look around rather than manipulating with the wand'', while some complained about the head movements: ``have to move my head a lot.''\replaced{, and some reported motion sickness.}{Many participants also reported experiencing motion sickness in it.}

\noindent\textbf{Curved map} was generally similar to the flat map. \replaced{Concerning}{Some commented on their feelings about the} curvature: ``it is like facing a curved TV''. \replaced{On}{Some were impressed by} visual design: ``It is the most visually impressive one.'' \replaced{Some perceived it as more distorted than flat map, some experienced motion sickness}{Yet, others reported it seemed more distorted than the flat map and felt more motion sickness with it than with the flat map}.
\vspace{-1.5em}
\section{Discussion}
\label{sec:discussion}
\vspace{-0.5em}
\noindent\textbf{Exocentric Globe:} 
Exocentric globes present geographic information without projection distortion, and this seems to greatly benefit accuracy for our tasks as described below. The overall response time is also comparatively good, except for one task (\emph{small variation} condition in area comparison). Overall, exocentric globe appears to be a good choice for the three fundamental geographic analysis tasks in VR with the following details and caveats:

\noindent\textit{Best for overall accuracy --}
The exocentric globe was the most accurate visualisation in almost all cases. One exception is in \emph{small variation} for area comparison, where it performed the worst. \\
\noindent\textit{Time for distance comparison similar to other visualisations --} The response time for the exocentric globe was similar to the other visualisations when comparing distances. \\
\noindent\textit{Area comparison slower than with maps --} Overall, participants were slower with the exocentric globe than the flat and curved maps. If we break the results down by task difficulties, however, performance for exocentric globe was similar to the flat and curved maps in \emph{easy} and \emph{far distance} conditions. However, it performed much slower than these two in the \emph{small variation} condition. We believe this is due to the relatively small variation of area between the two polygons, and the fact that participants can only view half of the globe. Participants had to rely on their memory to compare areas, and they tended to confirm their choice multiple times by using interactions to switch between viewing each of the two polygons. Interestingly, this did not happen in the \emph{small variation} condition in distance comparison tasks. A likely explanation is that comparing the magnitude of length is easier than area \cite{teghtsoonian1975psychophysics}, thus less interactions were needed in distance comparison tasks. \\
\noindent\textit{Fastest for direction estimation tasks --} Overall, the exocentric globe is the fastest visualisation for finding directions.

\vspace{-0.2em}
\noindent\textbf{Flat Map:} 
This visualisation is capable of presenting the entire surface of the world within the user's field of view, making it time-efficient for distance and area comparison.  However, projection and perceptual distortion appears to lead to poor accuracy in all direction estimation tasks, and the \emph{small variation} condition of distance and area comparison tasks. 

\noindent\textit{Relatively fast for distance and area comparison,} however, the accuracy is relatively low for distance comparison tasks. \\
\noindent\textit{Drop in accuracy for small variation --} In distance and area comparison tasks, accuracy was relatively high in \emph{easy} and \emph{far distance} conditions, but dropped significantly in the \emph{small variation} conditions. For distance comparison, no map projection can preserve all distances, thus, projection distortion makes comparing distances with small differences more difficult. For area comparison, although the map projection preserves the relative size of areas, the shape of the polygons is distorted (e.g., a polygon is elongated at the edge of a flat map), which affects area perception \cite{Teghtsoonian:1965gf}. \\
\noindent\textit{Poor for direction estimation --} The flat map was the least accurate visualisation and relatively slow in direction estimation.  
We attribute this to distortion effects.

\vspace{-0.2em}
\noindent\textbf{Egocentric Globe:}
This is the most immersive visualisation. It performed stably in accuracy across different difficulty conditions in the comparison of distances and areas. However, the perceptual distortion introduced by changing view point and the extra effort of body interaction (e.g.\ users needing to turn their head) make it a poor choice in VR for the three tasks tested.

\noindent\textit{Worst performance overall --} The egocentric globe performed significantly worse in almost all cases.  Particularly, for distance comparison, it seems less accurate and slower than other visualisations in \emph{easy} and \emph{far distance} conditions. \\
\noindent\textit{Good for small variation -- } Despite being the worst performer overall, one exception is in \emph{small variation} for distance comparison, in which, it seems to perform the best of all (both faster and more accurate).  One possible explanation is that it was the largest scale visualisation - i.e.\ it maximised the size of the distances relative to the participants' field of view at the default viewing distance.  While it is possible for participants to move closer to the visualisations to achieve a similar relative scale, doing so may be inconvenient or cost them time.\\
\noindent\textit{Stable across difficulties -- }  For distance and area comparison, the accuracy with egocentric globes stayed relatively stable across three difficulty conditions. \\
\noindent\textit{Motion sickness -- } Although we placed two fixed-position references (horizon lines) to help participants perceive their direction, they reported a relatively strong motion-sickness feeling in this visualisation. The egocentric spherical globe covered the full field of view all the time.  Together with the curvature of sphere, this seemed to cause participants to feel more motion-sickness. 

\vspace{-0.2em}
\noindent\textbf{Curved Map:}
This is generally an improvement on the flat map in VR. The curved map was more accurate than the flat map in direction estimation.  However, motion-sickness is a practical issue, but one that may be mitigated by shrinking the size of the map or by improved hardware.

\noindent\textit{Better than flat map for direction --}
Participants had greater accuracy with curved map than with flat map in direction estimation tasks, and performed similar to flat map in almost all other cases. \\
\noindent\textit{Accurate small variation area comparison --}
In area comparison tasks, curved map seems to be the most accurate visualisation in the \emph{small variation} condition, but was the least accurate one in \emph{far distance} condition. The possible reason might be that, the curved map is not an area-preserving visualisation, and the difference between distortions of areas are larger if two polygons are far apart. \\
\noindent\textit{Direction estimation --} 
In direction estimation tasks, the curved map outperformed the flat map in the \emph{close variation} condition. Less distortions of directions occurs in the curved map than the flat map. \\
\noindent\textit{Second worst for motion-sickness --}
Participants reported more feelings of motion-sickness than for exocentric globe and flat map. The reason might be similar as to the egocentric globe, the curved map covers 108$^\circ$ horizontally of the field of view, and the perception of curvature might produce extra motion-sickness.

\vspace{-0.2em}
\noindent\textbf{User interactions: }
In the egocentric globe, participants tended to interact significantly more than with other visualisations. This is also reflected in participants' feedback, and could be one reason for slow response times in the egocentric globe condition. 

The exocentric globe needed more controller interactions than curved and flat maps, and a similar degree of head movement. From Figure \ref{fig:interactions-result} and investigators' observation, it seems participants did not like to move themselves in space for all visualisations, especially in the exocentric globe, as its ratio of $\frac{controller\ interactions}{head\ interactions}$ is significantly larger than with other visualisations. This is possibly due to unfamiliarity with the VR environment, and the ease of manipulating visualisations using the controller.

\vspace{-1em}
\section{Conclusion and Future Work}
\label{sec:conclusion}

\begin{figure}
	\centering
	\includegraphics[width=.49\textwidth]{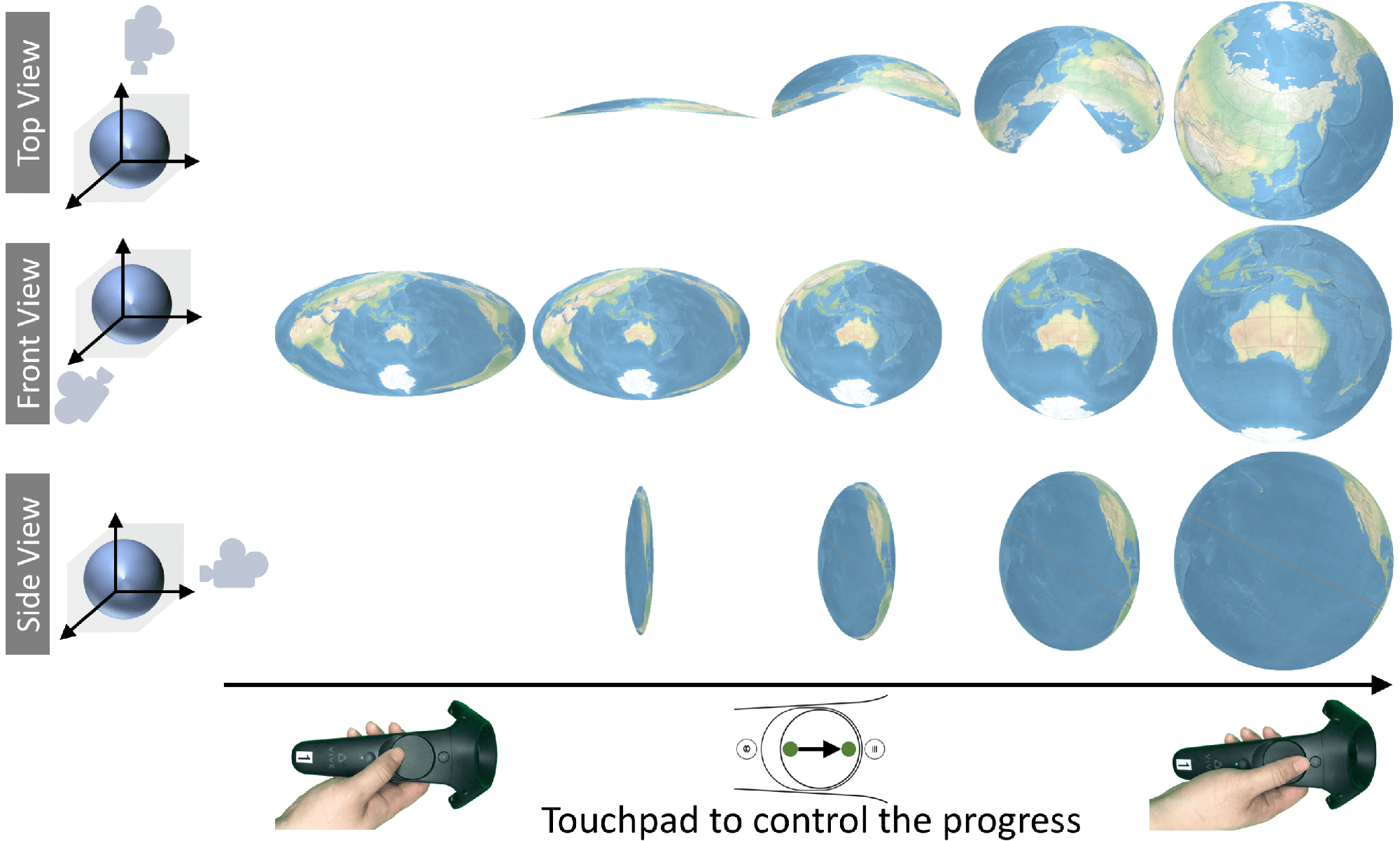}
	\vspace{-1em}
	\caption{Animated transformation from a flat map to an exocentric globe: top row - top view; second row - front view and bottom row - side view. The touchpad dynamically controls the progress.}
	\label{fig:morphing}
	\vspace{-2em}
\end{figure}

We have conducted the first user study evaluating the effectiveness of different visualisations of global  geography in VR. 
Of the four conditions and three task types tested, we found that the exocentric globe is generally the best choice of VR visualisation.  This is despite the fact that less of the earth's surface is visible in the exocentric globe than the other representations and that it has the most perceptual distortion, though no distortion due to map projection. We also found that the curved map had benefits over the flat map, but the curved map caused the users greater motion-sickness. In almost all cases the egocentric globe was found to be the least effective visualisation. 

\added{Our study was performed using VR HMDs as these offer a significantly better field of view compared to the currently available AR devices.  However, it is expected that AR technology will improve in this regard.
Thus, } our results have significant implications for the design of geovisualisation applications for VR\added{, AR and MR}, providing support for the use of exocentric globes when visualising data with global extent.
While our study found that the exocentric globe had the best overall performance of the four visualisations, the inability to show the entire surface hindered users in some tasks. We therefore think it would be reasonable to combine the exocentric globe with a map representation. As the curved map was more likely to cause motion sickness and the  flat map was more  familiar, we  are currently investigating how to combine the exocentric globe and flat map. 

As a first possible hybrid we have developed a prototype implementation that allows the viewer to interactively transition between exocentric globe and flat map. 
Due to the complexity of the transition, we allow the user to control the progress of the morphing. Fig.~\ref{fig:morphing} shows the animated transformation from a flat map to an exocentric globe, the reverse transformation  is symmetric. We used linear interpolation to transition between the  3D position of points in the rendered textures of the source visualisation and  the target visualisation. Refer to our video for a demonstration.  Evaluation of this hybrid visualisation remains future work.

Other future work is to design and evaluate other interactions including (non-physical) zooming. We also wish to investigate if our findings continue to hold for higher-level data analysis tasks of global spatial data using thematic maps such as  flow, choropleth or prism maps. We also believe that the curved map has considerable potential and wish to explore how we can reduce motion-sickness and also investigate the use of other map projections.

\noindent \textbf{Acknowledgments}\\
This research was supported under Australian Research Council's
Discovery Projects funding scheme (project number
DP180100755).  Data61, CSIRO (formerly NICTA) is funded by the Australian Government through the Department of Communications and the Australian Research Council through the ICT Centre for Excellence Program. We would like to thank all of our user study participants for their time and feedback. We would also like to thank the reviewers for their valuable comments.

\bibliographystyle{eg-alpha-doi}

\bibliography{template}

\end{document}